\documentclass[12pt]{article}

\usepackage{color}    
\usepackage[dvips]{graphicx}  
\usepackage{amsbsy}   
\usepackage{amsfonts}
\usepackage{amsmath}
\usepackage{amssymb}
\usepackage{color}
\usepackage{wasysym}
\usepackage{multirow}
\usepackage{mciteplus}
\usepackage{cite}
\usepackage{url}

\newcommand*{\ie}{\emph{i.e.}}

\newcommand{\kd}{\kappa^{D}}
\newcommand{\rd}{\rho^{D}}
\newcommand{\ku}{\kappa^{U}}
\newcommand{\ru}{\rho^{U}}
\newcommand{\kl}{\kappa^{L}}
\newcommand{\rl}{\rho^{L}}

\newcommand{\sba}{s_{\beta-\alpha}}
\newcommand{\cba}{c_{\beta-\alpha}}

\newcommand*{\Ave}[1]{\mathinner{\left\langle{#1}\right\rangle}}

\def\thefootnote{\fnsymbol{footnote}}

\begin{document}

\begin{center}
\Large\bf\boldmath
\vspace*{0.8cm} Flavor constraints on two-Higgs-doublet models\\ with general diagonal Yukawa couplings
\unboldmath
\end{center}

\vspace{0.8cm}
\begin{center}
F. Mahmoudi$^{1,}$\footnote{Electronic address: \tt mahmoudi@in2p3.fr} and O. St{\aa}l$^{2,}$\footnote{Electronic address: \tt oscar.stal@physics.uu.se}\\[0.4cm]
{\sl $^1$ Laboratoire de Physique Corpusculaire de Clermont-Ferrand (LPC), \\ Universit\'e Blaise Pascal, CNRS/IN2P3, 63177 Aubi\`ere Cedex, France }\\
{\sl $^2$ High-Energy Physics, Department of Physics and Astronomy,\\ 
Uppsala University, PO Box 516, 751 20 Uppsala, Sweden}\\
\end{center}

\vspace{1.0cm}
\begin{abstract}
\vspace*{0.2cm}
\noindent
We consider constraints from flavor physics on two-Higgs-doublet models (2HDM) with general, flavor-diagonal, Yukawa couplings. Analyzing the charged Higgs contribution to different observables, we find that $b\to s\gamma$ transitions and $\Delta M_{B_d}$ restrict the coupling $\lambda_{tt}$ of the top quark (corresponding to $\cot\beta$ in models with a $Z_2$ symmetry) to $|\lambda_{tt}|<1$ for $m_{H^+}\lesssim 500$ GeV. Stringent constraints from $B$ meson decays are obtained also on the other third generation couplings $\lambda_{bb}$ and $\lambda_{\tau\tau}$, but with stronger dependence on $m_{H^+}$. For the second generation, we obtain constraints on combinations of $\lambda_{ss}$, $\lambda_{cc}$, and $\lambda_{\mu\mu}$ from leptonic $K$ and $D_s$ decays.
The limits on the general couplings are translated to the common 2HDM types I -- IV with a $Z_2$ symmetry, and presented on the $(m_{H^+},\tan\beta)$ plane. The flavor constraints are most excluding in the type II model which lacks a decoupling limit in $\tan\beta$. We obtain a lower limit $m_{H^+}\gtrsim 300$ GeV in models of type II and III, while no lower bound on $m_{H^+}$ is found for types I and IV.
\end{abstract}
\newpage
\renewcommand{\thefootnote}{\arabic{footnote}}
\setcounter{footnote}{0}

\section{Introduction}
The LHC will hopefully shed some light soon on the mysteries of electroweak symmetry breaking. Meanwhile we are free to speculate whether nature is endowed with new physics beyond the minimal Higgs framework with one scalar doublet, as incorporated in the standard model (SM). One of the simplest -- and probably the most thoroughly studied -- extension of the Higgs sector is the two-Higgs-doublet model (2HDM), in which the SM Higgs sector is duplicated by introduction of another doublet of $\rm{SU}(2)_L$ \cite{Lee:1973iz,*Lee:1974jb, *Fayet:1974fj,*Fayet:1974pd,*Flores:1982pr}. The primary motivation for considering the 2HDM is often sought in supersymmetry \cite{Carena:2002es,*Djouadi:2005gj}, since the minimal supersymmetric standard model (MSSM) requires at least two Higgs doublets in order to cancel gauge anomalies and keep the superpotential analytic. This second requirement fixes the pattern of Yukawa couplings at tree-level. Many interesting results \cite{Gunion:1989we,Gunion:2002zf,Ginzburg:2004vp,Davidson:2005cw,*Haber:2006ue, Diaz:2002tp,*Branco:2005em,*Gunion:2005ja,*Maniatis:2006fs,*Ivanov:2006yq,*Ginzburg:2007jn,*Maniatis:2007vn,*Ivanov:2007de,*Sokolowska:2008bt,*Maniatis:2009vp,*Ferreira:2009wh,Gerard:2007kn,*deVisscher:2009zb} have also been obtained for more general two-Higgs-doublet models. These models can be thought of as effective theories for supersymmetric Higgs sectors beyond the MSSM (see e.g.~\cite{Dine:2007xi,*Carena:2009gx,*Antoniadis:2009rn} and references therein). In addition they provide a generic description for UV extensions of the SM that do not involve supersymmetry, but still contain two Higgs doublets in the low-energy effective theory.  Two-Higgs-doublet models of the  general form are the focus of the present work.

The additional scalar doublet in the 2HDM introduces new and interesting phenomenology, such as new sources of CP-violation, flavor-changing neutral currents (FCNC), the presence of a charged Higgs boson, and possibly even a dark matter candidate \cite{Ma:2006km}.
In the Yukawa sector, the additional doublet leads to a large number of couplings between fermions and scalars which are not fixed by the mass generation mechanism. When FCNCs are absent at tree-level, the most striking phenomenological difference between the SM and the 2HDM Yukawa sectors is the presence of a charged Higgs boson and its associated charged current. It is mainly through this interaction the 2HDM can reveal its presence in low-energy observables, such as the decay rates of $K$, $D$, and $B$ mesons, complementing searches for charged Higgs bosons at high-energy colliders. Results on the 2HDM Yukawa sector are therefore useful in both these contexts.

Constraints on the 2HDM parameters have been extensively studied in the literature, in particular for the type II model \cite{Cheung:2003pw,*WahabElKaffas:2007xd,*Flacher:2008zq,*Deschamps:2009rh,*Bona:2009cj} which is the form of the Yukawa sector in the MSSM at tree-level. For recent studies discussing flavor constraints on the charged Higgs boson in the MSSM, see \cite{Carena:2006ai,*Barenboim:2007sk,*Eriksson:2008cx}. To differentiate between the MSSM Higgs sector and a more general theory with two Higgs doublets, it is interesting to identify differences in phenomenologically viable signals, such as a charged Higgs boson contribution to flavor physics observables. Combining future LHC results with flavor physics could also be a means of testing the Yukawa couplings to more than one type and generation of fermions. 
To determine the reach in the flavor sector, we extend the previous results for the 2HDM type II to flavor constraints in the more general setting of arbitrary flavor-diagonal Yukawa couplings, \ie\ with no assumptions of coupling universality between fermions of different types or generations. Limits on the relevant couplings are evaluated for each observable separately. As a special case, we then translate the general results into constraints on the $(m_{H^+},\tan\beta)$ plane for types I -- IV \cite{Barger:1989fj} of the 2HDM with a $Z_2$ symmetry.\footnote{The phenomenology of the more unusual models type III and IV, which have different couplings for the down-type quarks and the leptons, has recently been studied in \cite{Aoki:2009ha,*Logan:2009uf,*Su:2009fz}.}

The organization of this paper is as follows: in Section~\ref{sec:theory} we give a brief introduction to the 2HDM, describing how the parameters of the potential are constrained by theoretical arguments and electroweak precision data. Following this discussion of the scalar sector, Section~\ref{sec:yukawa} deals with the 2HDM Yukawa sector in some detail. The different flavor observables of interest are introduced in Section~\ref{sec:flavor}, where also also our main results are presented as constraints on the individual Yukawa couplings from each observable separately. In Section~\ref{sec:combined} we specialize the general results of Section~\ref{sec:flavor} to models with a  Yukawa $Z_2$  symmetry. Finally, Section~\ref{sec:conclusions} presents the conclusions of this work.

\section{2HDM Fundamentals}
\label{sec:theory}
In a generic basis the most general, renormalizable, Higgs potential for two identical doublets $\Phi_1$ and $\Phi_2$, with hypercharge $Y=1$, is given by \cite{Gunion:2002zf}
\begin{equation}
  \begin{aligned}
    V_{\rm{2HDM}} = &\,m_{11}^2\Phi_1^\dagger\Phi_1+m_{22}^2\Phi_2^\dagger\Phi_2
    -\left[m_{12}^2\Phi_1^\dagger\Phi_2+\mathrm{h.c.}\right]
    \\
    &+\frac{1}{2}\lambda_1\left(\Phi_1^\dagger\Phi_1\right)^2
    +\frac{1}{2}\lambda_2\left(\Phi_2^\dagger\Phi_2\right)^2
    +\lambda_3\left(\Phi_1^\dagger\Phi_1\right)\left(\Phi_2^\dagger\Phi_2\right)
    +\lambda_4\left(\Phi_1^\dagger\Phi_2\right)\left(\Phi_2^\dagger\Phi_1\right)
    \\&+\left\{
    \frac{1}{2}\lambda_5\left(\Phi_1^\dagger\Phi_2\right)^2
    +\left[\lambda_6\left(\Phi_1^\dagger\Phi_1\right)
      +\lambda_7\left(\Phi_2^\dagger\Phi_2\right)
      \right]\left(\Phi_1^\dagger\Phi_2\right)
    +\mathrm{h.c.}\right\}.
  \end{aligned}
  \label{eq:pot_gen}
\end{equation}
In total there are $14$ free parameters in this model, including the complex phases which may be present in $\lambda_{5,6,7}$ and $m_{12}^2$. Restricting ourselves to the case without CP violation, all parameters are assumed to be real, and the number of free parameters is reduced to $10$. For the electroweak (EW) symmetry to be broken, the scalar mass matrix has at least one negative eigenvalue. At the minimum, $m_{11}^2$ and $m_{22}^2$ can be eliminated in favor of the vacuum expectation values $\Ave{\Phi_i}\equiv v_i/\sqrt{2}$. The 2HDM is invariant under a unitary transformation of the scalar doublets \cite{Ginzburg:2004vp,Davidson:2005cw,*Haber:2006ue}. A basis in the doublet space is chosen by specifying $\tan\beta\equiv v_2/v_1$. The overall scale (like in the SM) is set by $v^2=v_1^2+v_2^2\simeq(246\;\rm{GeV})^2$. For the sake of simplicity, we take $\lambda_6=\lambda_7=0$ in this section. This is required if the potential should respect a discrete $Z_2$ symmetry imposed on the $\Phi_i$ fields.\footnote{The $Z_2$ symmetry can still be (softly) broken by dimension two terms for non-zero values of $m_{12}^2$.} We will return to the importance of discrete symmetries later. The number of free parameters, $v$ being fixed, is down to six (plus $\tan\beta$ which specifies the basis).

When the EW symmetry is broken, three of the eight scalar degrees of freedom are used for the masses of the gauge bosons. The physical spectrum thus contains three neutral Higgs bosons and the charged Higgs pair $H^\pm$. Two of the neutrals are CP-even ($h$ and $H$, with $m_h<m_H$), and one ($A$) is CP-odd. The CP-even scalars mix with an angle $\alpha$. At the minimum, the original doublets are expanded according to
\begin{equation}
\Phi_1=\frac{1}{\sqrt{2}}\left(\begin{array}{c}
\displaystyle \sqrt{2}\left(G^+\cos\beta -H^+\sin\beta\right)  \\
\displaystyle v\cos\beta-h\sin\alpha+H\cos\alpha+\mathrm{i}\left( G^0\cos\beta-A\sin\beta \right)
\end{array}
\right)
\end{equation}
\begin{equation}
\Phi_2=\frac{1}{\sqrt{2}}\left(\begin{array}{c}
\displaystyle \sqrt{2}\left(G^+\sin\beta +H^+\cos\beta\right)  \\
\displaystyle v\sin\beta+h\cos\alpha+H\sin\alpha+\mathrm{i}\left( G^0\sin\beta+A\cos\beta \right)
\end{array}
\right).
\end{equation}
 Using the notation $s_\beta\equiv \sin\beta$, $c_\beta\equiv\cos\beta$, the tree-level Higgs masses are given in terms of the remaining potential parameters of Eq.~(\ref{eq:pot_gen}) by
\begin{equation}
m_A^2=\frac{m_{12}^2}{s_\beta c_\beta}-\lambda_5 v^2
\label{eq:mA}
\end{equation}
for the CP-odd Higgs,
\begin{equation}
m_{H,h}^2=\frac{1}{2}\left[M_{11}^2+M_{22}^2\pm\sqrt{\left(M_{11}^2-M^2_{22}\right)^2+4\left(M_{12}^2\right)^2}\right]
\end{equation}
for the CP-even Higgses, where the mass matrix is
\begin{equation}
M^2=m_A^2
\left(
\begin{array}{cc}
s_\beta^2 & -s_\beta c_\beta \\
-s_\beta c_\beta & c_\beta^2
\end{array}
\right)+v^2
\left(
\begin{array}{cc}
\lambda_1 c_\beta^2+\lambda_5 s_\beta^2 & (\lambda_3+\lambda_4)s_\beta c_\beta \\
(\lambda_3+\lambda_4)s_\beta c_\beta & \lambda_2 s_\beta^2+\lambda_5 c_\beta^2
\end{array}
\right),
\end{equation}
and finally
\begin{equation}
m_{H^+}^2=m_A^2+\frac{1}{2}v^2(\lambda_5-\lambda_4)
\label{eq:mHp}
\end{equation}
for the charged Higgs mass. Working at tree-level, the mass relations allow the input parameters $\lambda_{1-5}$ to be substituted by the four Higgs masses and the neutral sector mixing $\sin(\beta-\alpha)$, giving a set of physical input parameters.

The parameters in Eq.~(\ref{eq:pot_gen}) are bounded by the requirement that the potential is stable \cite{Deshpande:1977rw,*Sher:1988mj}, \ie\ that no direction in field space exists for which $V\to -\infty$. Necessary and sufficient conditions for stability are given by
\[
\lambda_1>0,\qquad \lambda_2>0,\qquad \lambda_3>-\sqrt{\lambda_1\lambda_2},
\]
\[
\lambda_3+\lambda_4-|\lambda_5|>-\sqrt{\lambda_1\lambda_2}.
\]
The parameters can also be bounded from above by arguments of perturbativity and perturbative unitarity of the $S$-matrix for Higgs and longitudinal vector boson scattering \cite{Huffel:1980sk,*Maalampi:1991fb,*Kanemura:1993hm,*Akeroyd:2000wc,*Ginzburg:2005dt}. We consider the upper limits $|\lambda_{i}|<4\pi$ (perturbativity) and $|L|<16\pi$ for the eigenvalues $L$ of the $S$-matrix. Assuming the 2HDM is perturbatively defined up to some scale $\Lambda$, these conditions should in principle be fulfilled at all scales $\mu<\Lambda$. The resulting limits on the $\lambda_i$ can then be studied in an RGE-improved treatment \cite{Nie:1998yn,*Kanemura:1999xf,*Ferreira:2009jb}. Since this is not the primary aim of this paper, we limit our discussion of the theoretical consistency to the input (EW) scale. When the potential has an \emph{exact} $Z_2$ symmetry ($m_{12}^2=0$), the requirement of stability, perturbativity, and unitarity leads to upper limits on the Higgs masses of $m_h\lesssim 600$ GeV, $m_H\lesssim 870$ GeV, $m_A\lesssim 870$ GeV, and $m_{H^+}\lesssim 780$ GeV. 
Breaking the $Z_2$ symmetry softly, the upper limits are further increased, as should be expected when introducing the new scale $m_{12}^2$. In the decoupling limit \cite{Gunion:2002zf}, the Higgs masses scale like $m_h\sim v$, and  $m_{H,A,H^+}\sim m_S\left(1+v^2/m_S^2\right)$, where $m_S$ is some high scale $m_S\gg v$ at which the heavy doublet is integrated out. The existence of the decoupling limit shows the absence of an absolute upper limit for the heavy Higgs masses in the general 2HDM. We should therefore focus on deriving \emph{lower} limits on the masses, which will require additional input from experiments.

\begin{figure}
\centering
\includegraphics[width=0.47\columnwidth]{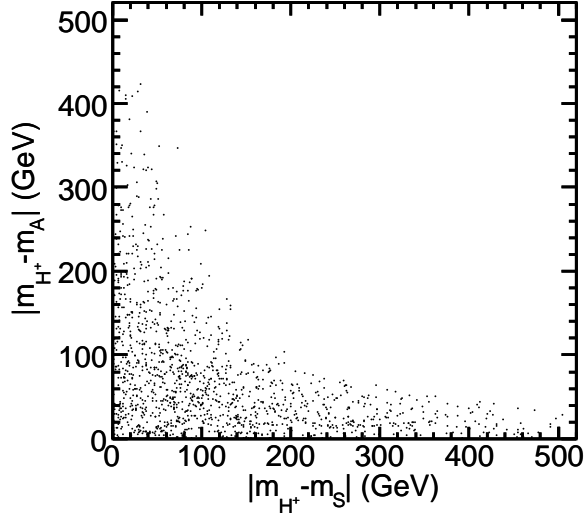}
\caption{Mass differences for 2HDM points which give contributions to $S$, $T$, and $U$ in agreement with the present limits \cite{Amsler:2008zzb} at the $2\,\sigma$ level.}
\label{fig:oblique}
\end{figure}

Before considering the Yukawa sector, we want to briefly discuss electroweak precision tests and the 2HDM contribution to the oblique parameters $S$, $T$, $U$ \cite{Peskin:1990zt,*Altarelli:1990zd,*Altarelli:1991fk,*Peskin:1991sw}. Fixing $m_h=114$~GeV, we perform a scan over the other 2HDM masses (with $m_{H_i}>m_h$) and $\sin(\beta-\alpha)$. For each point, the results of \cite{Grimus:2007if,*Grimus:2008nb} are used to evaluate the oblique parameters. Points which fulfill the PDG limits \cite{Amsler:2008zzb} at the $2\,\sigma$ level are illustrated in Fig.~\ref{fig:oblique}. The figure shows the mass splittings $|m_{H^+}-m_A|$ and $|m_{H^+}-m_S|$, where $m_S^2=m_H^2\sin^2(\beta-\alpha)+m_h^2\cos^2(\beta-\alpha)$ is a combined scalar mass. 
We see that both these mass splittings cannot be simultaneously $\mathcal{O}(v)$  for the 2HDM to be compatible with the EW precision tests. The limits $m_{H^+}\simeq m_A$ (independent of $\beta-\alpha$), $m_{H^+}\simeq m_{H}$ (with $\cos(\beta-\alpha)=0$), and $m_{H^+}\simeq m_h$ (with $\sin(\beta-\alpha)=0$) correspond to the known cases when custodial symmetry ensures $T\simeq 0$ \cite{Haber:1992py,*Pomarol:1993mu,Gerard:2007kn,*deVisscher:2009zb}.

\section{The Yukawa sector}
\label{sec:yukawa}
Assuming CP conservation, the general Yukawa Lagrangian of the 2HDM, expressed in the fermion mass eigenstates, is \cite{Davidson:2005cw,*Haber:2006ue}
\begin{equation}
-\mathcal{L}_Y=\overline{Q}_L\widetilde{\Phi}_1\eta_1^U U_R+\overline{Q}_L\Phi_1\eta_1^D D_R+\overline{Q}_L \Phi_1 \eta_1^L L_R+\overline{Q}_L\widetilde{\Phi}_2\eta_2^U U_R+\overline{Q}_L \Phi_2\eta_2^D D_R+\overline{Q}_L \Phi_2 \eta_2^L L_R,
\label{eq:genY}
\end{equation}
where $\widetilde{\Phi}_i\equiv \rm{i}\sigma_2\Phi_i^*$. The $\eta_i^F$, with $F=\{U,D,L\}$, are real $3\times 3$ Yukawa matrices, subject to the requirement that masses for the fermions are generated by the combination coupling to the vevs:
\begin{equation}
M^F=\frac{v}{\sqrt{2}}\left(\eta_1^F\cos\beta+\eta_2^F\sin\beta\right).
\label{eq:MQ}
\end{equation}
$M^F$ is here the (real and positive) mass matrix. It is convenient to introduce the short-hand notation
\begin{equation}
\kappa^F\equiv \eta_1^F\cos\beta+\eta_2^F\sin\beta.
\label{eq:kappa}
\end{equation}
 The combination orthogonal to $\kappa^F$, and which does not participate in giving mass to the fermions, is then
\begin{equation}
\begin{aligned}
\rho^F&\equiv-\eta_1^F\sin\beta+\eta_2^F\cos\beta .
\label{eq:rho}
\end{aligned}
\end{equation}
Performing a change of basis on the Higgs fields, we can also re-express the Yukawa Lagrangian, Eq.~(\ref{eq:genY}), in the basis $(H_1,H_2)$ where only one Higgs doublet (here taken to be $H_1$) acquires a vev. The resulting Lagrangian becomes
\begin{equation}
-\mathcal{L}_Y=\overline{Q}_L\widetilde{H}_1\ku U_R+\overline{Q}_LH_1\kd D_R+\overline{Q}_L H_1 \kl L_R+\overline{Q}_L\widetilde{H}_2\ru U_R+\overline{Q}_L H_2\rd D_R+\overline{Q}_L H_2 \rl L_R.
\end{equation}
$H_1$ (with vev $v_1=v$) generates the fermion masses and consequently couples  with $\kappa^F$, while $H_2$ has $v_2=0$ and couples with $\rho^F$. 

The Yukawa matrices for the Higgs doublets determine the couplings of the physical scalars to the fermions, which in an arbitrary basis are given by
\begin{equation}
  \begin{aligned}
    -\mathcal{L}_{\rm{Y}}&=\frac{1}{\sqrt{2}}\overline{D}\Bigl[\kd\sba+\rd\cba \Bigr]Dh
    +\frac{1}{\sqrt{2}}\overline{D}\Bigl[\kd\cba-\rd\sba \Bigr]DH+ \frac{\mathrm{i}}{\sqrt{2}}\overline{D}\gamma_5\rd DA \\
    &\,+\frac{1}{\sqrt{2}}\overline{U}\Bigl[\ku\sba+\ru\cba \Bigr]Uh
    +\frac{1}{\sqrt{2}}\overline{U}\Bigl[\ku\cba-\ru\sba \Bigr]UH- \frac{\mathrm{i}}{\sqrt{2}}\overline{U}\gamma_5\ru UA \\
    &\,+\frac{1}{\sqrt{2}}\overline{L}\Bigl[\kl\sba+\rl\cba \Bigr]Lh
    +\frac{1}{\sqrt{2}}\overline{L}\Bigl[\kl\cba-\rl\sba \Bigr]LH+ \frac{\mathrm{i}}{\sqrt{2}}\overline{L}\gamma_5\rl LA \\
    &\,+\frac{}{}\Bigl[\overline{U}\bigl(V_{\rm{CKM}} \rd P_R-\ru V_{\rm{CKM}} P_L\bigr)DH^+ + \overline{\nu}\rl P_RL H^+ + \rm{h.c.}\Bigr].  
\end{aligned}
\label{eq:yukawa}
\end{equation}
The general form of Eq.~(\ref{eq:yukawa}) exhibits tree-level FCNCs unless the $\rho^F$ are diagonal. A sufficient condition for this to occur is that each fermion type $F=\{D,U,L\}$ couples to only one Higgs doublet \cite{Glashow:1976nt}. In Eq.~(\ref{eq:genY}) this condition translates into demanding either $\eta^F_1=0$ or $\eta^F_2=0$, which through Eqs.~(\ref{eq:kappa}), (\ref{eq:rho}) leads to the relations $\rho^F=\kappa^F\cot\beta$ and $\rho^F=-\kappa^F\tan\beta$, respectively. A way to accomplish vanishing $\eta^F_i$ naturally is to impose a symmetry to prevent some of the couplings from appearing in Eq.~(\ref{eq:genY}), for example a $Z_2$ symmetry under which one Higgs doublet and some of the right-handed fermion fields are odd. Models with such a symmetry are called 2HDM \emph{types}. We choose a convention for the types I -- IV given by \cite{Barger:1989fj}.\footnote{The notation and meaning of the different types varies in the literature. Sometimes type Y (III) and type X (IV) is used. In supersymmetry, type III usually refers to the general model encountered when the $Z_2$ symmetry of the tree-level type II model is broken by higher order corrections. } The assignment of $Z_2$ charges, and the emergent coupling relations, are specified in Table~\ref{tab:Z2}.
\begin{table}
\centering
\begin{tabular*}{0.7\columnwidth}{@{\extracolsep{\fill}}ccccccc}
\hline
Type & $U_R$ & $D_R$ & $L_R$ & $\ru$ & $\rd$ & $\rl$ \\
\hline
I & $+$ & $+$ & $+$ & $\ku\cot\beta$ & $\kd\cot\beta$ & $\kl\cot\beta$ \\
II & $+$ & $-$ & $-$ & $\ku\cot\beta$ & $-\kd\tan\beta$ & $-\kl\tan\beta$ \\
III/Y & $+$ & $-$ & $+$ & $\ku\cot\beta$ & $-\kd\tan\beta$ & $\kl\cot\beta$ \\
IV/X & $+$ & $+$ & $-$ & $\ku\cot\beta$ & $\kd\cot\beta$ & $-\kl\tan\beta$ \\
\hline
\end{tabular*}
\caption{Assignment of $Z_2$ charges for the right-handed fermions, and the resulting relations among Yukawa coupling matrices in the $Z_2$-symmetric types of 2HDM Yukawa sectors. The Higgs doublets have $Z_2$ quantum numbers  $-$ $(\Phi_1)$ and $+$ $(\Phi_2)$.}
\label{tab:Z2}
\end{table}

The $Z_2$-symmetric 2HDM types constitute minimal models for the Yukawa sector in the sense that they are one-parameter descriptions. From a phenomenological viewpoint, a rather natural step beyond the types are models with Yukawa alignment \cite{Pich:2009sp}. Assuming that the Yukawa matrices for each type of fermions are proportional by a constant $\xi^F$, one can define three $\tan\beta$-like parameters through $\rho^F=\tan\beta^F\kappa^F$, where $\tan\beta^F=\frac{\xi^F-\tan\beta}{1+\xi^F\tan\beta}$. Since $\kappa^F$ and $\rho^F$ can be diagonalized simultaneously, the model is free of tree-level FCNCs. A further generalization is offered by models where the off-diagonal Yukawa elements are non-zero, but still sufficiently small to avoid the experimental bounds from FCNC. Using a a Hermitian six-zero texture for the Yukawa matrices,\footnote{The phenomenology of the related case with four-zero textures was recently discussed in \cite{DiazCruz:2004tr,*DiazCruz:2004pj,*DiazCruz:2009ek}.} a phenomenologically viable ansatz for this type of models was introduced by Cheng and Sher \cite{Cheng:1987rs}
\begin{equation}
[\rho^F]_{ij}=[\lambda^F]_{ij}\frac{\sqrt{2m_im_j}}{v}\qquad\rm{(no\ sum)}.
\label{eq:ChengSher}
\end{equation}
It leads to a natural suppression of the off-diagonal elements for $\lambda\sim\mathcal{O}(1)$ by the quark mass hierarchy, in particular for the first generation where the experimental constraints are most restrictive. Different scenarios have been investigated in \cite{Atwood:1996vj}. The Cheng-Sher ansatz provides a parametrization useful also for general, flavor-diagonal, Yukawa couplings. From the definition of $\kappa^F$, it follows that the diagonal elements are
\begin{equation}
[\rho^F]_{ii}=[\lambda^F]_{ii}\frac{\sqrt{2}\,m_{i}}{v}=[\lambda^F]_{ii}[\kappa^F]_{ii} \qquad\rm{(no\ sum)}.
\label{eq:diagonal}
\end{equation}
This relation describes general flavor-diagonal couplings with a preserved mass hierarchy.\footnote{From the appearance of equation~\eqref{eq:diagonal} one could refer to this case as \emph{generalized alignment}, although the motivation in terms of a proportionality between the Yukawa matrices no longer holds.} The nine $\lambda^F_{ii}$ (one for each fermion) are free $\tan\beta$-like parameters. Both the case of alignment ($\lambda^U_{11}=\lambda_{22}^U=\lambda_{33}^U=\tan\beta^U$, similarly for $D$ and $L$), and the models with a $Z_2$ symmetry (e.g. type II: $\lambda^U_{ii}=\cot\beta$, $\lambda_{ii}^D=\lambda_{ii}^F=-\tan\beta$) are contained as special cases of this general framework. We shall use this phenomenological approach in the following. The Yukawa couplings have been implemented in the generic code 2HDMC \cite{Eriksson:2009ws} used throughout this work. When there is no risk of confusion, we use the notation $\lambda_{tt}\equiv\lambda^U_{33}$, $\lambda_{bb}\equiv\lambda^D_{33}$, etc. Inspired by the $Z_2$-symmetric models, we further restrict the couplings to, at most, allow for hierarchies of the order
\begin{equation}
\begin{aligned}
& m_d|\lambda_{dd}|\sim m_u|\lambda_{uu}|\sim 0\\
& m_s|\lambda_{ss}|\sim m_c|\lambda_{cc}|,\qquad
m_c|\lambda_{cc}|\sim m_b|\lambda_{bb}|,\qquad
m_b|\lambda_{bb}|\sim m_t|\lambda_{tt}|\\
& m_e|\lambda_{ee}| \ll m_{\mu}|\lambda_{\mu\mu}| \ll m_{\tau}|\lambda_{\tau\tau}| .
\end{aligned}
\end{equation}
This means couplings to the first generation are always neglected. Couplings to the nearest neighbors in mass can have a hierarchy allowing them to be of similar magnitude, hence they must be considered simultaneously in observables where both couplings appear. No strong hierarchy is allowed between generations of equal charge fermions, since this would be unnatural under the Cheng-Sher assumption.
%The $Z_2$ symmetric 2HDM types I -- IV thus correspond to special cases where a) universality relations of the sort $\lambda_{uu}=\lambda_{cc}=\lambda_{tt}$ hold  for all classes of fermions and b) couplings to different fermion types are related either by $\lambda_{tt}=\lambda_{bb}$ or $\lambda_{tt}=-1/\lambda_{bb}$.
%We take a more general attitude and investigate which couplings are constrained by each flavor observable separately. In Sec.~\ref{sec:combined} we make universality assumptions to connect to the more well-known results for the $Z_2$ symmetric models, in particular those for type II.

%As our model for the 2HDM Yukawa sector we take a general Cheng-Sher ansatz, as given by Eq.~(\ref{eq:ChengSher}), but with the additional restriction that off-diagonal elements can be kept under control to the degree that their contribution to any observable is negligible.  Effectively we are therefore left with Eq.~(\ref{eq:diagonal}): one free coupling $\lambda_{ii}$ for each fermion.  
\section{Constraints from flavor physics}
\label{sec:flavor}
The most stringent experimental constraints on the 2HDM come from flavor physics. When the Yukawa couplings are diagonal in flavor space, one could still have sizeable contributions to many observables due to charged Higgs exchange, either at tree-level or at one-loop. We shall focus on such processes here. The Yukawa coupling dependence from the lowest order charged Higgs contribution to the interesting observables is listed in Table~\ref{tab:yukdep}, together with the current experimental results and the SM  expectations. For the numerical evaluation we use SuperIso~v2.6 \cite{Mahmoudi:2007vz,*Mahmoudi:2008tp}, both for the SM and the 2HDM contributions. The slight differences in some of the SM values compared to earlier published results are explained by parametric updates. The top quark pole mass is set to the latest combined Tevatron value $m_t=173.1\pm 0.6\pm 1.1$ \cite{Tev:2009ec}, corresponding to $\overline{m}_t=164.0$~GeV using the 2-loop relation. For the other quark masses, we use the PDG $\overline{\mathrm{MS}}$ values \cite{Amsler:2008zzb} $\overline{m}_b=4.20^{+0.17}_{-0.07}$~GeV, $\overline{m}_c=1.27^{+0.07}_{-0.11}$~GeV, and $m_s=104^{+26}_{-34}$~MeV  as input. When calculating leptonic- and semi-leptonic decays of mesons, the running quark masses appearing in the $H^+$ couplings are evaluated at the scale of the meson mass. This ensures that the effective four-fermion operators $\overline{u}_Ld_R\overline{\ell}_R\nu_L$, obtained when integrating out $H^+$, renormalize in the same way as the corresponding quark currents \cite{Carena:1999py}. Constraints on the 2HDM parameters from each observable are presented as exclusion regions at $95\%$  C.L., using Gaussian two-sided confidence intervals and one degree of freedom.
\begin{table}
\begin{tabular*}{\textwidth}{@{\extracolsep{\fill}}llll}
\hline
Observable & Couplings & Experimental value &  Standard Model\\
\hline
$\rm{BR}(B\to X_s\gamma$) & $\lambda_{tt}^2$, $\lambda_{tt}\lambda_{bb}$ & $(3.52\pm 0.23\pm 0.09)\times 10^{-4}$  \cite{Barberio:2008fa} & $(3.07\pm 0.22)\times 10^{-4}$ \\
$\Delta_0(B\to K^*\gamma$) & $\lambda_{tt}^2$, $\lambda_{tt}\lambda_{bb}$ & $(3.1\pm 2.3)\times 10^{-2}$ \cite{Mahmoudi:2007vz,*Mahmoudi:2008tp} & $(7.8\pm 1.7 )\times 10^{-2}$\\
$\Delta M_{B_d}$ & $\lambda_{tt}^2$ & $(0.507\pm 0.004)$ ps$^{-1}$ \cite{Barberio:2008fa} & $(0.53\pm 0.08)$ ps$^{-1}$\\
$\rm{BR}(B_u\to \tau\nu_\tau$) & $\lambda_{bb}\lambda_{\tau\tau}$ & $(1.73\pm 0.35)\times 10^{-4}$ \cite{Charles:2004jd,*Tisserand:2009ja} & $(0.95\pm 0.27)\times 10^{-4}$ \\
$\xi_{D\ell\nu}$ & $\lambda_{bb}\lambda_{\tau\tau}, \lambda_{cc}\lambda_{\tau\tau}$ & $0.416\pm 0.117\pm 0.052$ \cite{Aubert:2007dsa} & $0.30\pm 0.02$ \\
$R_{\ell 23}(K\to \mu\nu_\mu$) & $\lambda_{ss}\lambda_{\mu\mu}$ & $1.004\pm 0.007$  \cite{Antonelli:2008jg} & $1$  \\
$\rm{BR}(D_s\to \mu\nu_\mu$) & $\lambda_{ss}\lambda_{\mu\mu}$, $\lambda_{cc}\lambda_{\mu\mu}$ & $(5.8\pm 0.4)\times 10^{-3}$ \cite{Akeroyd:2009tn} & $(4.98\pm 0.15)\times 10^{-3}$\\
$\rm{BR}(D_s\to \tau\nu_\tau$) & $\lambda_{ss}\lambda_{\tau\tau}$, $\lambda_{cc}\lambda_{\tau\tau}$ & $(5.7\pm 0.4)\times 10^{-2}$ \cite{Akeroyd:2009tn} & $(4.82\pm 0.14)\times 10^{-2}$ \\
\hline
\end{tabular*}
\caption{List of observables sensitive to a charged Higgs contribution at lowest order, their dependence on the 2HDM Yukawa couplings, the current experimental and SM values. For cases where the experimental value is a combination of several values, we reference the paper where the combination is performed. The SM values are evaluated with SuperIso~v2.6 \cite{Mahmoudi:2007vz,*Mahmoudi:2008tp}, using the methods and input parameters described in the text.}
\label{tab:yukdep}
\end{table}

Negative searches for the charged Higgs at LEP lead to a lower limit $m_{H^+}\gtrsim 78$ -- $90$ GeV \cite{Abdallah:2003wd}. The exact value depends on the $H^+$ decay mode. Strictly speaking, these limits are only valid under the assumption that the $H^+$ decays with total $\mathrm{BR}=1$ in the modes $H^+\to\tau^+\nu_\tau$ and $H^+\to c\bar{s}$, with certain benchmarks investigated for $H^+\to W^{+(*)}A$ showing similar exclusion. Even if the 2HDM with general Yukawa couplings could certainly allow for exotic decay channels, hierarchical couplings still favor decays into these channels. We will therefore respect the LEP limits in the following and not consider the possibility $m_{H^+}\lesssim m_W$.

The first flavor observable we consider is the branching ratio for the rare FCNC inclusive decay $B\to X_s\gamma$, which receives contributions from the charged Higgs at the same level as the $W$ contribution in the SM. Using an effective Hamiltonian approach, the leading order 2HDM contribution to the relevant Wilson coefficients at the matching scale $\mu_W$ is \cite{Hou:1987kf,*Grinstein:1990tj}
\begin{equation}
\delta C_{7,8}^{\rm{2HDM}}(\mu_W)=\frac{1}{3}\lambda_{tt}^2F_{7,8}^{(1)}(y)-\lambda_{tt}\lambda_{bb}F_{7,8}^{(2)}(y)
\label{eq:C7}
\end{equation}
where $y=\overline{m}_t^2/m_{H^+}^2$, and
\begin{equation}
\begin{aligned}
F_7^{(1)}(y)&=\frac{y(7-5y-8y^2)}{24(y-1)^3}+\frac{y^2(3y-2)}{4(y-1)^4}\ln y\\
F_8^{(1)}(y)&=\frac{y(2+5y-y^2)}{8(y-1)^3}-\frac{3y^2}{4(y-1)^4}\ln y\\
F_7^{(2)}(y)&=\frac{y(3-5y)}{12(y-1)^2}+\frac{y(3y-2)}{6(y-1)^3}\ln y\\
F_8^{(2)}(y)&=\frac{y(3-y)}{4(y-1)^2}-\frac{y}{2(y-1)^3}\ln y.
\end{aligned}
\end{equation}
The lowest order expression given by Eq.~(\ref{eq:C7}) captures the essential dependence on the Yukawa couplings $\lambda_{tt},\lambda_{bb}$ even when higher order corrections are included. The numerical evaluation of $\mathrm{BR}(B\to X_s\gamma)$ \cite{Mahmoudi:2007vz,*Mahmoudi:2008tp} is performed to NNLO for the SM \cite{Misiak:2006ab}, and to NLO for the 2HDM contribution \cite{Ciuchini:1997xe}.
\begin{figure}
\centering
\includegraphics[width=0.47\columnwidth]{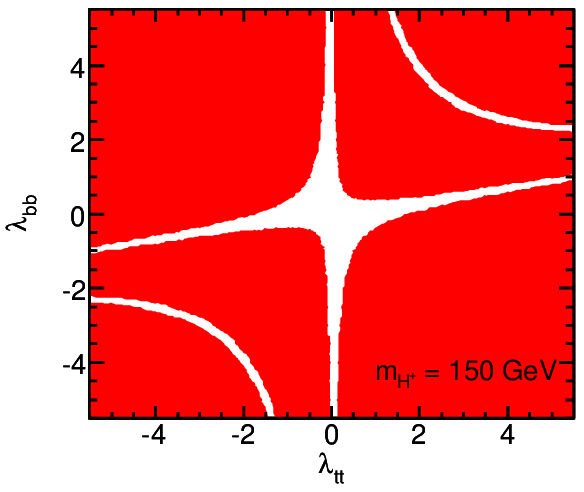}
\includegraphics[width=0.47\columnwidth]{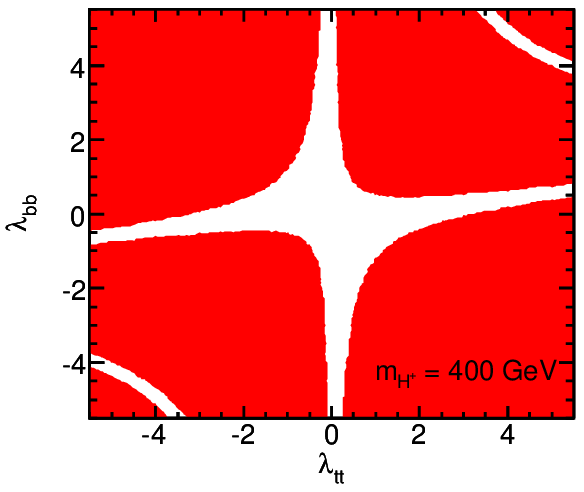}
\caption{Constraints on $(\lambda_{tt}, \lambda_{bb})$ from $\mathrm{BR}(B\to X_s\gamma)$ for fixed $m_{H^+}=150$~GeV (left) and $m_{H^+}=400$~GeV (right). The highlighted region is excluded at 95\% C.L.}
\label{fig:bsgamma}
\end{figure}
Results on the excluded regions of parameter space $(\lambda_{tt},\lambda_{bb})$ are shown in Fig.~\ref{fig:bsgamma} for two values of the charged Higgs mass: $m_{H^+}=150$~GeV and $m_{H^+}=400$~GeV. These were chosen having the well-known limit $m_{H^+}\gtrsim 300$~GeV for the 2HDM II in mind. When $m_{H^+}=150$~GeV, we find that a large fraction of the parameter space is excluded, with a preference for smaller values of the couplings in the allowed region. We also see allowed regions in Fig.~\ref{fig:bsgamma} with simultaneously large and equal sign values for $\lambda_{bb}$ and $\lambda_{tt}$. For these fine-tuned values, a cancellation occurs in the new physics contribution to the Wilson coefficients. The situation is similar for $m_{H^+}=400$ GeV, except that the size of the allowed region increases. As expected from Eq.~(\ref{eq:C7}), taking $\lambda_{bb}\to 0$ is in general not enough to avoid generating a large $H^+$ contribution, but $\lambda_{tt}\to 0$ always is.

Another quantity measured in $b\to s\gamma$ transitions is the degree of isospin asymmetry in the exclusive decay mode $B\to K^*\gamma$, defined as \cite{Kagan:2001zk,*Bosch:2001gv}
\begin{equation}
\Delta_{0-}\equiv\frac{\Gamma(\bar{B}^0\to \bar{K}^{*0}\gamma)-\Gamma(\bar{B}^-\to \bar{K}^{*-}\gamma)}{\Gamma(\bar{B}^0\to \bar{K}^{*0}\gamma)+\Gamma(\bar{B}^-\to \bar{K}^{*-}\gamma)}.
\end{equation}
 This observable gives more stringent constraints on the model parameters than $\mathrm{BR}(B\to X_s\gamma)$ in several MSSM scenarios \cite{Ahmady:2006yr,*Mahmoudi:2007gd}. Using the NLO prediction we investigate, for the first time, constraints from $\Delta_{0-}$ on the Yukawa sector of the general 2HDM.
\begin{figure}
\centering
\includegraphics[width=0.47\columnwidth]{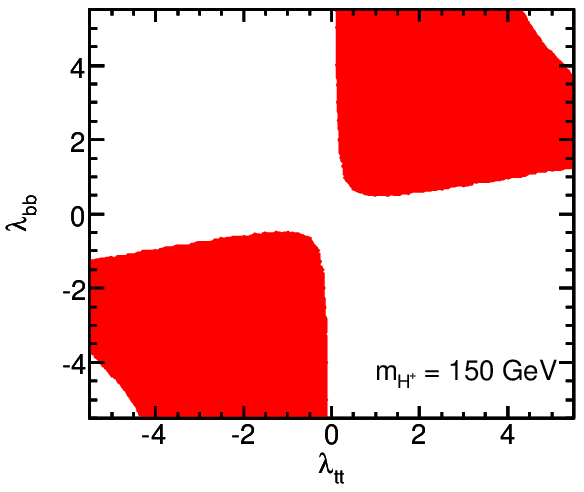}
\includegraphics[width=0.47\columnwidth]{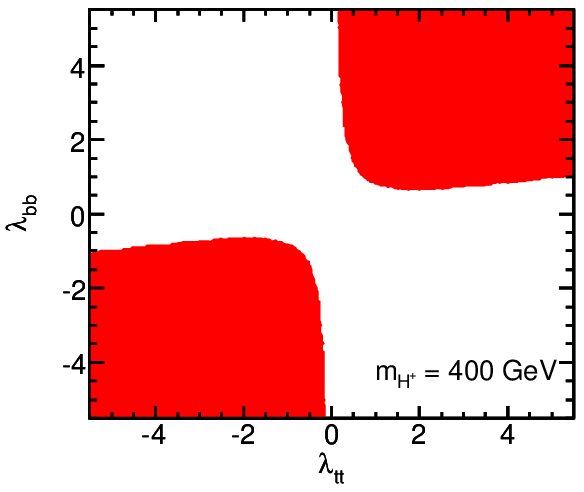}
\caption{Constraints on $(\lambda_{tt}$, $\lambda_{bb})$ from the isospin asymmetry $\Delta_{0-}$ for fixed $m_{H^+}=150$~GeV (left) and $m_{H^+}=400$~GeV (right). The highlighted region is excluded at 95\% C.L.}
\label{fig:delta0}
\end{figure}
The excluded region in $(\lambda_{tt},\lambda_{bb})$ is presented in Fig.~\ref{fig:delta0}. To facilitate a comparison with the results  for the branching ratio, the same values as in Fig.~\ref{fig:bsgamma} are chosen for $m_{H^+}$. From Figs.~\ref{fig:bsgamma} and \ref{fig:delta0}, one notices a similarity in the regions excluded by $\mathrm{BR}(B\to X_s\gamma)$ and $\Delta_{0-}$. This results from the common dependence of both observables on the Wilson coefficient $C_7$: $\mathrm{BR}(B\to X_s\gamma)$ is proportional to $C_7^2$, while $\Delta_{0-}$ varies like $C_7^{-1}$ to first order. The isospin asymmetry results exclude the large-coupling solution observed in Figure~\ref{fig:bsgamma} for same-sign couplings.

The $\Delta M_{B_d}$ and $\Delta M_{B_s}$ mass differences measured in $B^0$--$\bar{B}^0$ mixing are sensitive to charged Higgs exchange through box diagrams involving internal top quarks. Including the leading order contributions, the 2HDM prediction for $\Delta M_{B_d}$ is \cite{Geng:1988bq,Ball:2006xx}
\begin{equation}
\Delta M_{B_d}=\frac{G_F^2m_t^2f_{B_d}^2\hat{B}_dM_B|V_{td}^*V_{tb}|^2\eta_b}{24\pi^2}\Biggl[I_{WW}(y^W)+I_{WH}(y^W,y^H,x)+I_{HH}(y^H)\Biggr],
\label{eq:DeltaMb}
\end{equation}
where $y^i=\overline{m}_t^2/m_i^2$ ($i=W,H^+$), $x=m_{H^+}^2/m_W^2$, and
\begin{equation}
\begin{aligned}
I_{WW}&=1+\frac{9}{1-y^W}-\frac{6}{(1-y^W)^2}-\frac{6}{y^W}\left(\frac{y^W}{1-y^W}\right)^3\ln y^W\\
I_{WH}&=\lambda_{tt}^2\, y^H\left[\frac{(2x-8)\ln y^H}{(1-x)(1-y^H)^2}+\frac{6x\ln y^W}{(1-x)(1-y^W)^2}-\frac{8-2y^W}{(1-y^W)(1-y^H)}\right]\\
I_{HH}&=\lambda_{tt}^4\, y^{H}\left[\frac{1+y^H}{(1-y^H)^2}+\frac{2y^H\ln y^H}{(1-y^H)^3} \right].
\end{aligned}
\end{equation}
Approximate effects of short-distance QCD corrections are incorporated in Eq.~(\ref{eq:DeltaMb}) through the factor $\eta_b=0.552$ \cite{Buchalla:1995vs}.
The non-perturbative decay constant $f_{B_d}$ and the bag parameter $\hat{B}_d$ are evaluated simultaneously from lattice QCD. We use the value  $f_{B_d}\hat{B}_d^{1/2}=216\pm 15$~MeV \cite{Gamiz:2009ku}, adding all errors in quadrature. Since the uncertainty in $f_{B_d}\hat{B}_d^{1/2}$ is correlated to the error in the corresponding parameters for $B_s$ decays, and since the theoretical uncertainty dominates the experimental uncertainties, $\Delta M_{B_d}$ and $\Delta M_{B_s}$ do not provide independent constraints on the 2HDM. We therefore consider only $\Delta M_{B_d}$ which has the smallest total uncertainty of the two.
\begin{figure}
\centering
\includegraphics[width=0.47\columnwidth]{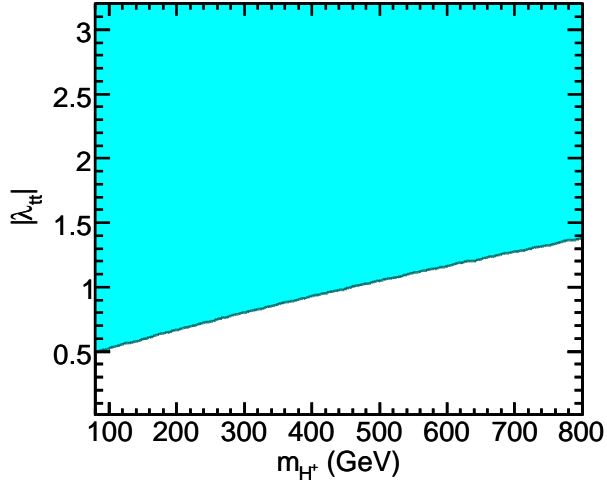}
\caption{Constraints on $\left(m_{H^+},|\lambda_{tt}|\right)$ from $\Delta M_{B_d}$. The highlighted region is excluded at 95\% C.L.}
\label{fig:DeltaMB}
\end{figure}
In Fig.~\ref{fig:DeltaMB} the excluded region for $|\lambda_{tt}|$ is shown as a function of $m_{H^+}$. Nothing can be said about the sign of $\lambda_{tt}$ since it enters $\Delta M_{B_d}$ quadratically. From Fig.~\ref{fig:DeltaMB} it is clear that large values of $|\lambda_{tt}|$ are disfavored; we find that $|\lambda_{tt}|>1$ is ruled out for $m_{H^+}\lesssim 500$ GeV, independently of the other couplings.

There are a number of decays  which the charged Higgs can mediate already at tree-level. 
The first such observable we consider is $B_u\to \tau\nu_\tau$, which has a small branching ratio in the SM caused by helicity suppression. Since no suppression occurs for the $H^+$, the two contributions may be of similar magnitude, leading to sizable interference and a reduced branching fraction. The 2HDM contribution factorizes, giving \cite{Hou:1992sy,*Akeroyd:2003zr,*Akeroyd:2007eh}
\begin{equation}
\mathrm{BR}(B_u\to \tau\nu_\tau)=
\frac{G_F^2f_B^2|V_{ub}|^2}{8\pi\Gamma_B}m_Bm_\tau^2\left(1-\frac{m_\tau^2}{m_B^2}\right)^2R_{\tau\nu}^{\rm{2HDM}},
\end{equation}
with
\begin{equation}
R_{B\tau\nu}^{\rm{2HDM}}=\left[1-\left(\frac{m_B^2}{m_{H^+}^2}\right)\lambda_{bb}\lambda_{\tau\tau}\right]^2.
\label{eq:RBtaunu}
\end{equation}
For the $B$ decay constant we use the value  $f_B=190\pm 13$~MeV \cite{Gamiz:2009ku}, which together with $V_{ub}$ constitutes the main theoretical uncertainty. We use the combined CKMfitter value $|V_{ub}|=3.87\pm 0.09\pm 0.46$ \cite{Charles:2004jd,*Tisserand:2009ja}.
\begin{figure}
\centering
\includegraphics[width=0.47\columnwidth]{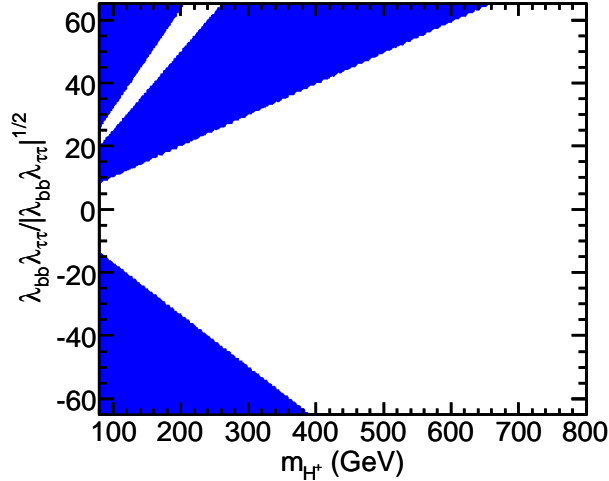}
\caption{Constraints on $\left(m_{H^+},\lambda_{bb}\lambda_{\tau\tau}\right)$ from $\mathrm{BR}(B_u\to\tau\nu_\tau)$. The highlighted region is excluded at 95\% C.L. The y-axis normalization corresponds to $\tan\beta$ in the 2HDM II (for positive values only).}
\label{fig:Btaunu}
\end{figure}
Since only the product $\lambda_{bb}\lambda_{\tau\tau}$ enters Eq.~(\ref{eq:RBtaunu}), the exclusion region is presented for this quantity versus $m_{H^+}$. The result is shown in Fig.~\ref{fig:Btaunu}, where $\lambda_{bb}\lambda_{\tau\tau}/|\lambda_{bb}\lambda_{\tau\tau}|^{1/2}$ on the y-axis corresponds to $\tan\beta$ in the familiar type II model (for positive values). $B_u\to\tau\nu_\tau$ excludes a large region of the parameter space, with stronger limits obtained on the product of couplings for lower values of $m_{H^+}$. Around $m_{H^+}=100$--$250$ GeV, there is a region for $\lambda_{bb}\lambda_{\tau\tau}>20$ where the interference is exactly twice the SM amplitude, but opposite in sign, leading to a cancellation in Eq.~(\ref{eq:RBtaunu}).

Another third generation observable, complementary to $B_u\to\tau\nu_\tau$, is the semi-leptonic mode $B\to D\tau\nu_\tau$. With the 2HDM contribution included, the differential decay rate with respect to $w=v_B\cdot v_D$ is given by \cite{Grzadkowski:1991kb,*Kiers:1997zt,*Kamenik:2008tj,Nierste:2008qe}
\begin{equation}
\frac{\mathrm{d}\Gamma(B\to D\tau \nu_\tau)}{\mathrm{d}w} = \frac{G_F^2|V_{cb}|^2 m_B^5}{192\pi^3}\rho_V(w) \times\left[1 - \frac{m_{\tau}^2}{m_B^2}\, \left\vert 1-t(w)\, \frac{m_b\lambda_{bb}-m_c\lambda_{cc}}{(m_b-m_c)m^2_{H^{+}}}\lambda_{\tau\tau} \right\vert^2 \rho_S(w) \right],
\label{eq:BDtaunu}
\end{equation}
where $\rho_V$ ($\rho_S$) are vector (scalar) form factors and $t(w)=m_B^2+m_D^2-2wm_Dm_B$. It has been shown \cite{Nierste:2008qe} that comparing differential distributions directly would be a superior method to extract the charged Higgs contribution. However, since the collected statistics is still too low, this method has so far not been pursued experimentally. To reduce the uncertainty from the vector form factor, we consider the ratio \cite{Grzadkowski:1991kb,*Kiers:1997zt,*Kamenik:2008tj}
\begin{equation}
\xi_{D\ell\nu}=\frac{\mathrm{BR}(B\to D\tau\nu_\tau)}{\mathrm{BR}(B\to De\nu_e)},
\end{equation}
where the 2HDM contributes only to the numerator.
\begin{figure}
\centering
\includegraphics[width=0.47\columnwidth]{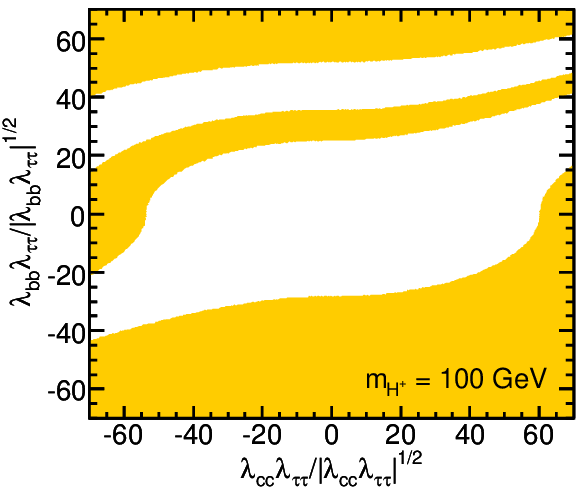}
\includegraphics[width=0.47\columnwidth]{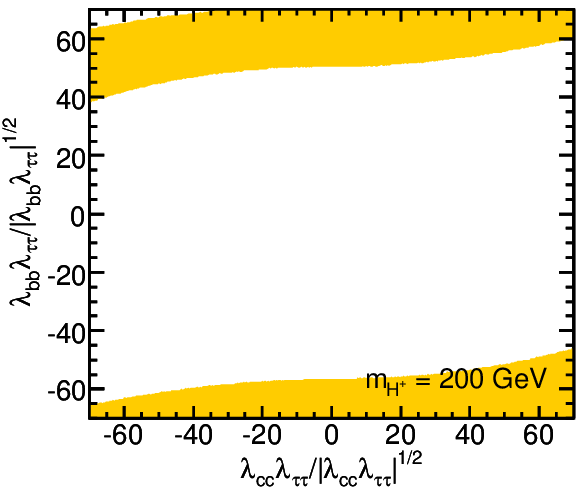}
\caption{Constraints on $\left(\lambda_{cc}\lambda_{\tau\tau}, \lambda_{bb}\lambda_{\tau\tau}\right)$ from $\mathrm{BR}(B \to D\tau\nu_\tau)$ for fixed $m_{H^+}=100$~GeV (left) and $m_{H^+}=200$~GeV (right). The highlighted region is excluded at 95\% C.L. The axes are normalized such that the values correspond to $\tan\beta$ $(\cot\beta)$ in the $Z_2$ symmetric versions of the 2HDM.}
\label{fig:BDtaunu}
\end{figure}
The resulting constraints based on $\xi_{D\ell\nu}$ are shown in Fig.~\ref{fig:BDtaunu} for $m_{H^+}=100$, $200$ GeV. Similarly to the result for $B_u\to\tau\nu_\tau$, the 2HDM contribution in Eq.~(\ref{eq:BDtaunu}) can be twice the SM contribution with opposite sign, leading to the two disjoint exclusion regions as observed in Fig.~\ref{fig:BDtaunu}. When $\lambda_{cc}\ll\lambda_{bb}$ -- like in the 2HDM II at high $\tan\beta$ -- the effective constraint on $\lambda_{bb}\lambda_{\tau\tau}$ can be combined with that from $B_u\to\tau\nu_\tau$ to cover fully the cancellation region observed for low $m_{H^+}$.

The last $B$ decay we consider is $B_s\to\mu^+\mu^-$, which has so far not been observed experimentally. The SM prediction for the branching ratio is
\begin{equation}
\mathrm{BR}(B_s \to \mu^+ \mu^-)_\mathrm{SM}=(3.2\pm 0.5) \times 10^{-9},
\end{equation}
while the current experimental limit, derived by the CDF collaboration, is \cite{Aaltonen:2007kv}:
\begin{equation}
\mathrm{BR}(B_s \to \mu^+ \mu^-) < 5.8 \times 10^{-8}
\end{equation}
at 95\% C.L. The gap between the SM prediction and the current experimental limit makes this observable particularly interesting in SUSY, since this difference leaves room for SUSY contributions. In the 2HDM however, we found that the experimental limit can be reached only for very large values of $\lambda_{\mu\mu}$ and $\lambda_{bb}$ or $\lambda_{tt}$ and small charged Higgs masses. The constraining power of $B_s\to\mu^+\mu^-$  in this study is hence rather limited as compared to the other flavor observables and therefore we do not investigate it further. 

Recently it was suggested \cite{Antonelli:2008jg} to use the observable
\begin{equation}
R_{\ell 23}\equiv\left| \frac{V_{us}(K_{\ell 2})}{V_{us}(K_{\ell 3})} \times \frac{V_{ud}(0^+ \to 0^+)}{V_{ud}(\pi_{\ell 2})} \right|
\end{equation}
from  $K\to\mu\nu_\mu$ transitions to search for charged Higgs contributions. In this expression $V_{us}(K_{\ell i})$ means $V_{us}$ measured in leptonic kaon decay with $i$ particles in the final state (two leptons and $i-2$ neutral pions). The same notation applies for $\pi_{\ell 2}$ while $V_{ud}(0^+\to 0^+)$ refers to $V_{ud}$ measured in nuclear beta decay. The 2HDM contribution to $R_{\ell 23}$ is given by 
\begin{equation}
R_{\ell 23}=\left|1-\frac{m^2_{K^+}}{M^2_{H^+}}\left(1 - \frac{m_d}{m_s}\right)\lambda_{ss}\lambda_{\mu\mu}\right|,
\end{equation}
where we use $m_d/m_s=1/20$ \cite{Amsler:2008zzb}. The major source of uncertainty in extracting $R_{\ell 23}$ experimentally originates in the necessary input of the form factor ratio $f_K/f_\pi$ from lattice QCD. From an unquenched calculation with staggered fermions, the HPQCD and UKQCD groups have determined $f_K/f_\pi=1.189\pm 0.007$ \cite{Follana:2007uv} which is the number adopted in \cite{Antonelli:2008jg}. We will make use of that same result, keeping in mind that alternative determinations with larger errors exist. Using these would of course reduce the constraining power of $K\to \mu\nu_\mu$ transitions.
\begin{figure}
\centering
\includegraphics[width=0.47\columnwidth]{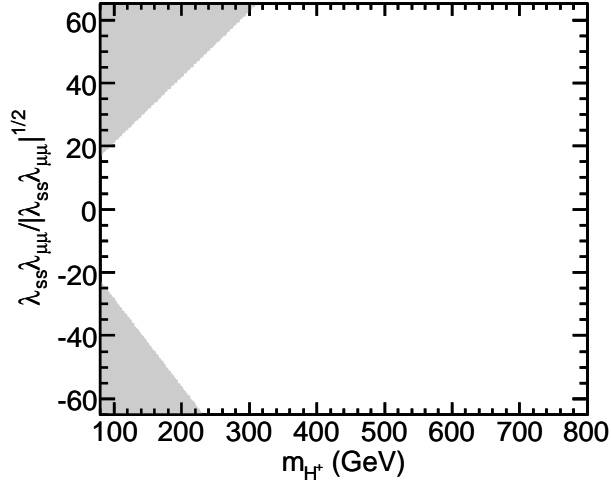}
\caption{Constraints on $\left(m_{H^+},\lambda_{ss}\lambda_{\mu\mu}\right)$ from $R_{\ell 23}(K\to \mu\nu_\mu)$. The highlighted region is excluded at 95\% C.L. The axis normalization is similar to $B_u\to\tau\nu_\tau$.}
\label{fig:Kmunu}
\end{figure}
Fig.~\ref{fig:Kmunu} shows the result for the product $\lambda_{ss}\lambda_{\mu\mu}$  as a function of $m_{H^+}$. Exclusion is obtained for low $m_{H^+}$ and large absolute values of the coupling, $|\lambda_{ss}\lambda_{\mu\mu}|\gtrsim 20$ for $m_{H^+}=100$~GeV. The $R_{\ell 23}$ observable is interesting since it provides constraints on the Yukawa couplings to the second generation of fermions. 

The modes $D_s \to \mu\nu_\mu$ and $D_s\to\tau\nu_\tau$ were shown in \cite{Akeroyd:2009tn} to give constraints on $(m_{H^+},\tan\beta)$ in 2HDM II competitive to those from $B$ meson decays. Here we would like to stress the additional role played by these observables as tests of the couplings to the second generation of quarks in connection with both second and third generation leptons. At tree-level, the $D_s\to\ell\nu_\ell$ branching ratio is given by \cite{Akeroyd:2003jb,Dobrescu:2008er}
\begin{equation}
\mathrm{BR}(D_s\to\ell\nu_\ell)=\frac{G_F^2f_{D_s}^2|V_{cs}|^2 M_{D_s}}{8\pi\Gamma_{D_s}}m_{\ell}^2\left(1-\frac{m_{\ell}^2}{m_{D_s}^2}\right)^2R_{D\ell\nu},
\label{eq:Dslnu}
\end{equation}
with the 2HDM factor
\begin{equation}
R_{D\ell\nu}=\Biggl[1-m_{D_s}^2\frac{m_s\lambda_{ss}-m_c\lambda_{cc}}{(m_c+m_s)m_{H^+}^2}\lambda_{\ell\ell}\Biggr]^2 .
\label{eq:RDslnu}
\end{equation}
From lattice QCD we adopt the value $f_{D_s}=241\pm 3$ MeV \cite{Follana:2007uv} which has the smallest quoted uncertainty. A major source of concern in Eq.~(\ref{eq:RDslnu}) is the dependence on the light quark masses. Effects of parametric uncertainties in the masses on the constraints from $D_s$ was investigated in \cite{Akeroyd:2009tn}. We will not repeat this discussion here, but treat the center values as exact quantities.
\begin{figure}
\centering
\includegraphics[width=0.47\columnwidth]{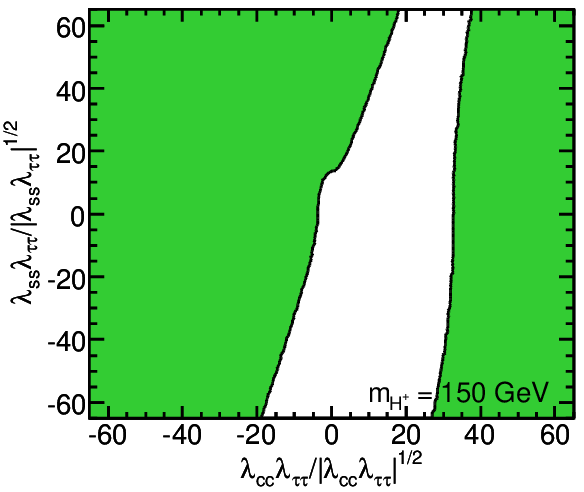}
\includegraphics[width=0.47\columnwidth]{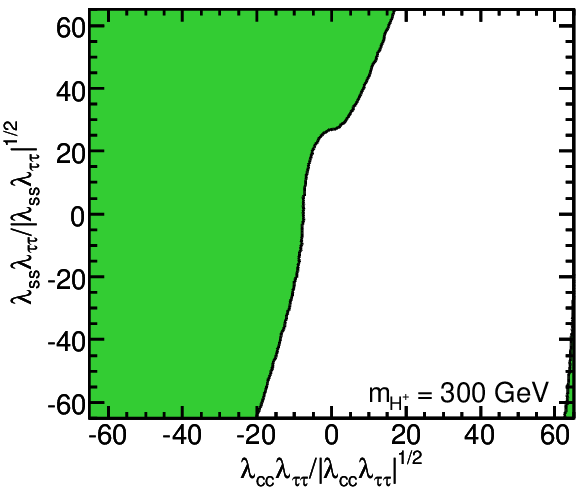}
\caption{Constraints on $\left(\lambda_{cc}\lambda_{\tau\tau}, \lambda_{ss}\lambda_{\tau\tau}\right)$ from $\mathrm{BR}(D_s \to \tau\nu_\tau)$ for fixed $m_{H^+}=150$~GeV (left) and $m_{H^+}=300$~GeV (right). The highlighted region is excluded at 95\% C.L. The axes are normalized similarly to Fig.~\ref{fig:BDtaunu}.}
\label{fig:Dstaunu}
\end{figure}
\begin{figure}
\centering
\includegraphics[width=0.47\columnwidth]{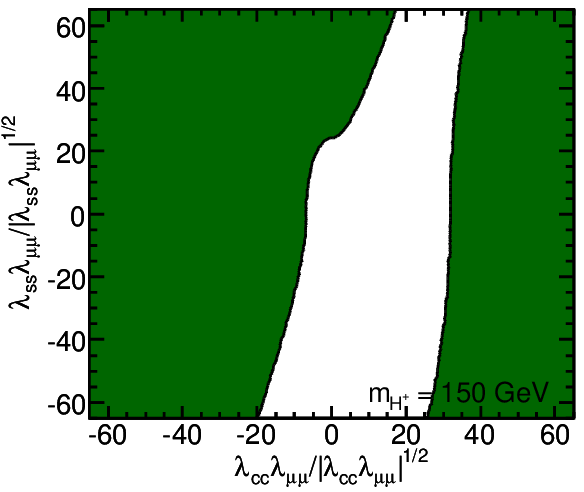}
\includegraphics[width=0.47\columnwidth]{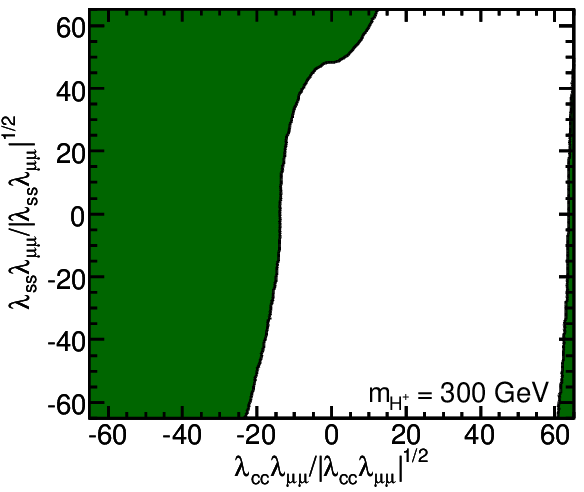}
\caption{Constraints on $\left(\lambda_{cc}\lambda_{\mu\mu}, \lambda_{ss}\lambda_{\mu\mu}\right)$ from $\mathrm{BR}(D_s \to \mu\nu_\mu)$ for fixed $m_{H^+}=150$~GeV (left) and $m_{H^+}=300$~GeV (right). The highlighted region is excluded at 95\% C.L. The axes are normalized similarly to Fig.~\ref{fig:BDtaunu}.}
\label{fig:Dsmunu}
\end{figure}
Constraints on the 2HDM couplings from the two modes $D_s\to\tau\nu_\tau$ and $D_s\to\mu\nu_\mu$ are summarized in Figs.~\ref{fig:Dstaunu} and \ref{fig:Dsmunu}, respectively. The excluded regions appear quite similar in these figures, as a result of the helicity suppression in the SM, which gives the same relative 2HDM contribution to both decay modes. This is interesting for the prospects of testing the hypothesis $\lambda_{\tau\tau}=\lambda_{\mu\mu}$. A similar test could be performed using the third generation quarks by considering $B_u\to\mu\nu_\mu$ in addition to $B_u\to\tau\nu_\tau$. Using the same input as above, the SM prediction becomes $\mathrm{BR}(B_u\to\mu\nu_\mu)=4.3\times 10^{-7}$.  Unfortunately, experiments are not yet sensitive at this level, with the current best limit $\mathrm{BR}(B_u\to\mu\nu_\mu)<1.6\times 10^{-6}$ \cite{Aubert:2008ri} reported at $95\%$ C.L.

For $D_s\to\mu\nu_\mu$, one may also combine the result with that from $K\to\mu\nu_\mu$ on the coupling $\lambda_{ss}\lambda_{\mu\mu}$, for which we found  $-40<\lambda_{ss}\lambda_{\mu\mu}<30$ when $m_{H^+}=150$ GeV. For $m_{H^+}=300$ GeV we obtained essentially no constraint from $K\to\mu\nu_\mu$.

\section{Models with a Yukawa $Z_2$ symmetry}
\begin{figure}
\centering
\includegraphics[width=0.48\columnwidth]{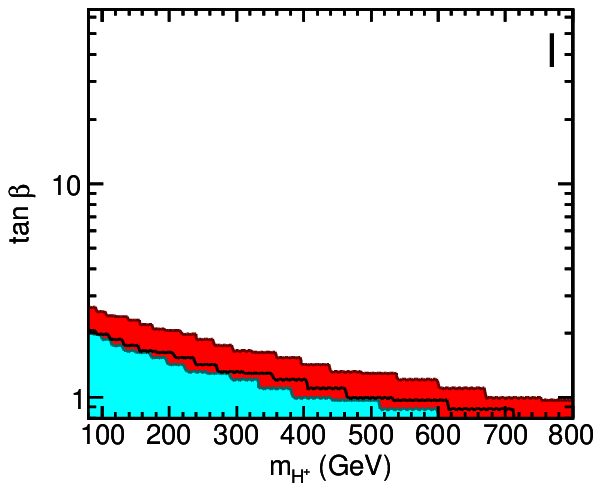}
\includegraphics[width=0.48\columnwidth]{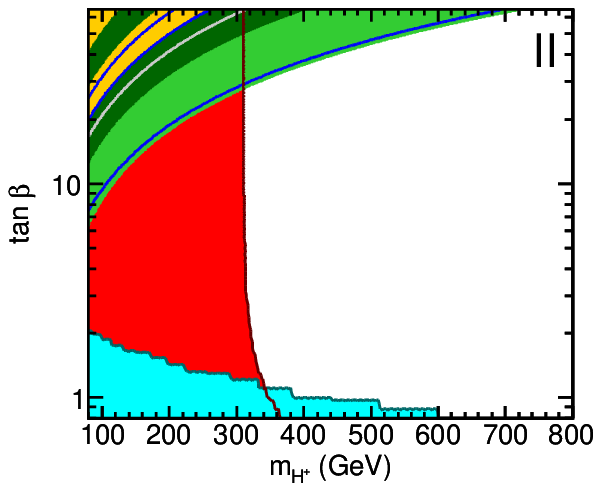}
\includegraphics[width=0.48\columnwidth]{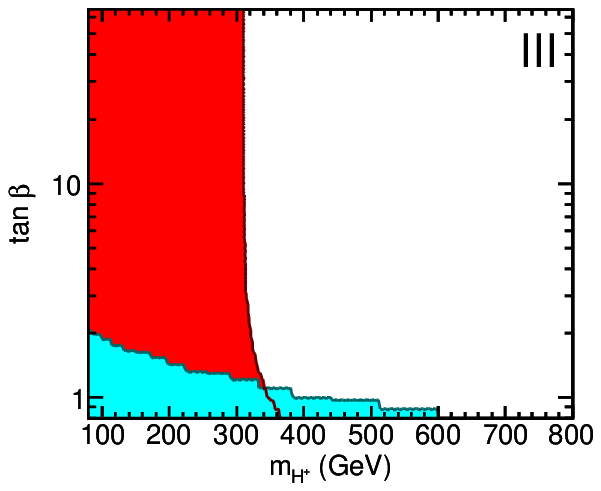}
\includegraphics[width=0.48\columnwidth]{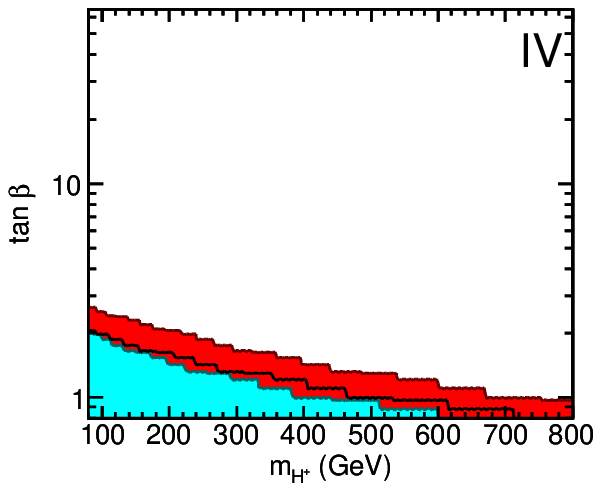}
\caption{Excluded regions of the ($m_{H^+},\tan\beta$) parameter space for $Z_2$-symmetric 2HDM types. The color coding is as follows: $\mathrm{BR}(B\to X_s\gamma)$ (red), $\Delta_{0-}$ (black contour), $\Delta M_{B_d}$ (cyan), $B_u\to\tau\nu_\tau$ (blue), $B\to D\tau\nu_\tau$ (yellow), $K\to\mu\nu_\mu$ (gray contour), $D_s\to\tau\nu_\tau$ (light green), and $D_s\to\mu\nu_\mu$ (dark green).}
\label{fig:combined}
\end{figure}

Having discussed constraints on the general model in the previous section, we now specialize to the 2HDM with a Yukawa $Z_2$ symmetry. In the language of the general model, this corresponds to universal couplings over the fermion generations and types with $\tan\beta$ the only free parameter, e.g.~$1/\lambda_{tt}=1/\lambda_{cc}=-\lambda_{bb}=-\lambda_{ss}=\tan\beta$ for type II couplings. The coupling patterns in the four different models we consider are described in Table~\ref{tab:Z2}.
The results, as shown in Fig.~\ref{fig:combined} in the ($m_{H^+},\tan\beta$) plane, are superimposed to combine the exclusion regions of all flavor observables in the same figure. This also means that some regions of the parameter space are excluded by more than one observable, such as the high $\tan\beta$ region in the type II model.
\label{sec:combined}

From Fig.~\ref{fig:combined}, we first note the exclusion of low $\tan\beta< 1$ in all four models for $m_{H^+}<500$~GeV. This exclusion comes as a result of three observables: $\mathrm{BR}(B\to X_s\gamma)$, $\Delta_0$, and $\Delta M_{B_d}$. The constraints at low $\tan\beta$ are similar between the models, since the couplings to the up-type quarks are universal. In the type I model, a value of $\tan\beta>1$ signals decoupling of one Higgs doublet from the whole fermion sector. In 2HDM types II and III, which share the same coupling pattern for the quarks, there exists a $\tan\beta$-independent lower limit of $m_{H^+}\gtrsim 300$ GeV imposed by $\mathrm{BR}(B\to X_s\gamma)$. No generic lower limit on $m_{H^+}$ is found in type I and type IV models. Constraints for high $\tan\beta$ are only obtained in the type II model. The reason behind this is that the leptonic and semi-leptonic observables require $\tan\beta$-enhanced couplings $\lambda_{dd}\lambda_{\ell\ell}\sim \tan^2\beta\gg 1$ ($d=d,s,b$) for the contributions to be interesting. In the 2HDM III, and IV these couplings are instead always $\lambda_{dd}\lambda_{\ell\ell}=-1$, while in type I they are proportional to $\cot^2\beta$.

\begin{figure}
\centering
\includegraphics[width=0.3\columnwidth]{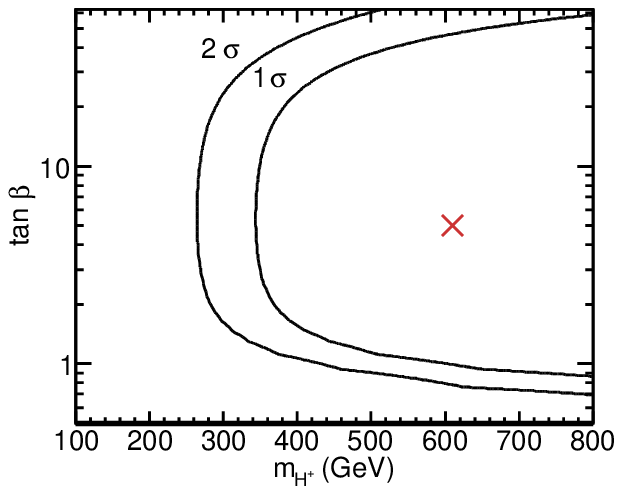}
\includegraphics[width=0.3\columnwidth]{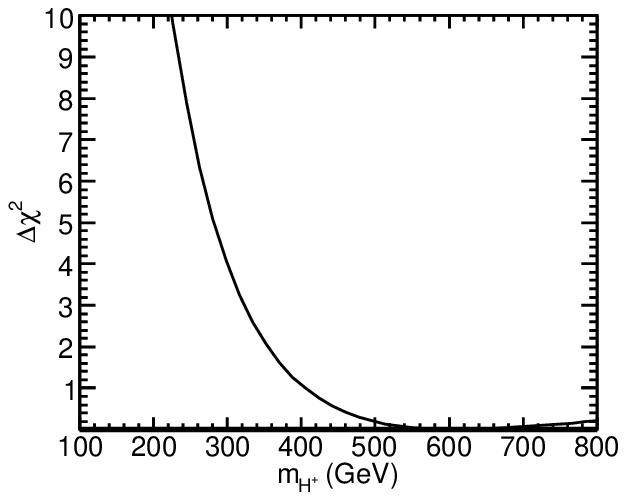}
\includegraphics[width=0.3\columnwidth]{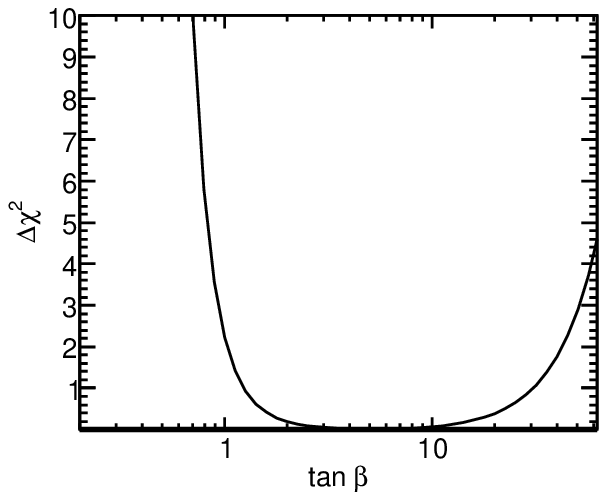}
\caption{Combined parameter estimation in ($m_{H^+},\tan\beta$) for the 2HDM II (left). The red cross indicates the best fit point $m_{H^+}=609$~GeV, $\tan\beta=5$. The contours show levels of $\Delta\chi^2=2.30$ $(6.18)$, corresponding to probabilities for $1\,\sigma$ $(2\,\sigma)$ Gaussian confidence intervals with two degrees of freedom. The center and right panels show variation of $\Delta\chi^2$ when changing the parameters $m_{H^+}$ and $\tan\beta$ around their best fit values.}
\label{fig:bestfit}
\end{figure}

An alternative approach to constraining the parameter space is to assume the 2HDM is the correct theory and estimate the parameters by fitting to the available data. To compare with our results in Figure~\ref{fig:combined}, we perform a basic $\chi^2$ fit  to the flavor data of the parameters ($m_{H^+},\tan\beta$) for the $Z_2$ symmetric models. The $\chi^2$ measure is constructed in the usual way
$
\chi^2=\sum_i\frac{\left(\mathcal{O}^{\rm{exp}}_i-\mathcal{O}_i^{\rm{2HDM}}\right)^2}{\sigma_i^2},
$
where $\sigma_i^2=(\sigma_i^{\rm{exp}})^2+(\sigma_i^{\rm{2HDM}})^2$ combines  experimental and theoretical uncertainties in quadrature. Parametric uncertainties  are not included.
We use the eight observables listed in Table~\ref{tab:chi2_type2} as independent probes of the 2HDM. Minimizing the $\chi^2$, the first observation is that the 2HDM types I, III, and IV, do not contain enough predictive power for the flavor observables to restrict the parameter space ($m_{H^+},\tan\beta$). In all these models, the decoupling limit in one variable (or both) is essentially the best fit.
For the 2HDM II on the other hand, we obtain a best fit point: $m_{H^+}=609$ GeV, $\tan\beta=5$. The fit is illustrated in Fig.~\ref{fig:bestfit}. The obtained exclusion region is similar to the combination presented in Fig.~\ref{fig:combined}. At the level of $2\,\sigma$, charged Higgs masses $m_{H^+}<260$ GeV are excluded ($m_{H^+}<300$~GeV for one degree of freedom). The exclusion in $\tan\beta$ extends to $\tan\beta<1$ and $\tan\beta>60$ for $m_{H^+}=500$ GeV. Both the lower and the upper limits on $\tan\beta$ become stricter for lower $m_{H^+}$.
In Table~\ref{tab:chi2_type2}, the results for the 2HDM II are given in more detail. The results for $b\to s\gamma$ transitions is what drives the fit away from decoupling in $m_{H^+}$. In fact, \emph{all} other measurements favor new physics contributions of opposite sign to those generated by the 2HDM II. This results in a $\tan\beta$ close to its value of minimum influence, $\tan\beta=\sqrt{m_t/m_b}\sim 6.5$ ($\tan\beta=\sqrt{m_c/m_s}\sim 3.5$) for the third (second) generation, since no decoupling limit in $\tan\beta$ exists in the 2HDM II. Figure~\ref{fig:bestfit} shows that the $\chi^2$ does not deteriorate significantly when increasing $m_{H^+}$, illustrating that decoupling is not far from being the best fit to the 2HDM II either. The reduced $\chi^2/\mathrm{d.o.f.}=2.1$ for the 2HDM fit, versus $\chi^2/\mathrm{d.o.f.}=1.8$ for the SM reference, confirms that the flavor data does not favor a 2HDM II over the SM at this point.
\begin{table}
\centering
\begin{tabular}{lllclcr@{.}l}
\hline
Observable & Experimental & SM prediction & $\Delta\chi^2_{\rm{SM}}$ & 2HDM fit & $\Delta\chi^2_{\rm{2HDM}}$& \multicolumn{2}{c}{Pull}\\
\hline
$\rm{BR}(B\to X_s\gamma$)     & $3.52\times 10^{-4}$  & $3.07\times 10^{-4}$ & $1.65$& $3.59\times 10^{-4}$  & $0.04$ & $0$  & $21$ \\
$\Delta_0(B\to K^*\gamma$)    & $3.1\times 10^{-2}$   & $7.8\times 10^{-2}$  & $2.82$ & $7.0\times 10^{-2}$   & $1.88$ & $1$  & $37$\\
$\Delta M_{B_d}$ (ps$^{-1}$)  & $0.507$  & $0.53$ & $0.08$ & $0.53$  & $0.10$ & $0$  & $32$\\
$\rm{BR}(B_u\to \tau\nu_\tau$)& $1.73\times 10^{-4}$  & $0.95\times 10^{-4}$ & $1.71$ & $0.95\times 10^{-4}$  & $1.72$ & $-1$& $31$\\
$\xi_{D\ell\nu}$              & $0.416$               & $0.30$              & $0.84$ & $0.30$               & $0.84$ & $-0$ & $91$\\
$R_{\ell 23}(K\to \mu\nu_\mu$)& $1.004$               & $1.000$              & $0.33$ & $1.000$               & $0.33$ & $-0$ & $58$ \\
$\rm{BR}(D_s\to \mu\nu_\mu$)  & $5.8\times 10^{-3}$   & $4.98\times 10^{-3}$ & $3.32$ & $4.98\times 10^{-3}$  & $3.36$ & $-1$ & $83$\\
$\rm{BR}(D_s\to \tau\nu_\tau$)& $5.7\times 10^{-2}$   & $4.82\times 10^{-2}$ & $3.82$ & $4.82\times 10^{-2}$  & $3.82$ & $-1$ & $95$\\
\hline
                              &                       & Total $\chi^2(\nu)$: & $14.6\,(8)$     & &  $12.1\,(6)$ &\multicolumn{2}{c}{} \\
\hline
\end{tabular}
\caption{Best fit of the 2HDM II to the flavor observables compared to the results for the SM. The experimental and theoretical uncertainties are given in Table~\ref{tab:yukdep}.}
\label{tab:chi2_type2}
\end{table}

\section{Conclusions}
\label{sec:conclusions}
Allowing for non-universal Yukawa couplings, the 2HDM can modify significantly the theoretical predictions for many flavor physics observables. Using the public codes 2HDMC \cite{Eriksson:2009ws} and SuperIso \cite{Mahmoudi:2007vz,*Mahmoudi:2008tp}, we have exploited this fact and used available experimental results to place constraints on the couplings of the second and third generation fermions to the charged Higgs boson in models with general, flavor-diagonal, Yukawa couplings. Strong and universal constraints are found on the top quark coupling $\lambda_{tt}$ from $\Delta M_B$ and $b\to s\gamma$ transitions, requiring $|\lambda_{tt}|\lesssim 1$ for $m_{H^+}\lesssim 500$~GeV. Observables such as $B_u\to\tau\nu_\tau$ or $D_s\to\ell\nu_\ell$, with a tree-level charged Higgs contribution depending on products of quark and lepton couplings, show sensitivity at low masses $m_{H^+}\lesssim 300$~GeV both for the second and the third generations. Typically, the exclusion limit starts at values $|\lambda_{qq}\lambda_{\ell\ell}|^{1/2}\gtrsim 10$ for $m_{H^+}=100$~GeV with softer limits for higher masses.

For the versions of the 2HDM with a Yukawa $Z_2$ symmetry, the lower limit $m_{H^+}\gtrsim 300$ GeV from $\mathrm{BR}(B\to X_s\gamma)$ is valid for type II and III models, while no general limit on $m_{H^+}$ exists for types I and IV. The bound on $\lambda_{tt}$ in the general model translates into a requirement of $\tan\beta>1$. The tree-level observables investigated yield constraints almost exclusively on the type II model at high $\tan\beta$. For this model, we find that only the observables from $b\to s\gamma$ transitions favor a non-zero 2HDM contribution, leading to a best fit point $m_{H^+}= 609$~GeV, $\tan\beta= 5$ close to decoupling.

The region of 2HDM parameter space where the LHC experiments expect sensitivity (small $m_{H^+}$, large $\tan\beta$) \cite{deRoeck:942733,Aad:2009wy} is very interesting for most of the observables discussed here. Should a charged Higgs be discovered at the LHC, the underlying model still needs to be determined. A low mass charged Higgs points to either the existence of additional low-energy states (as in supersymmetry), or non-type II couplings for $H^+$. The 2HDM Yukawa sector is accessible through flavor physics. In favorable cases, coupling universality between $\lambda_{\tau\tau}$ and $\lambda_{\mu\mu}$ can be tested using leptonic $B$ and $D_s$ decays which are helicity suppressed in the SM. Likewise it might be possible to test universality over the fermion generations by combining results from $B$ decays with those from $K\to\mu\nu$ or $D_s\to\ell\nu$. To summarize, a charged Higgs boson discovery would provide great opportunities for a rich interplay between flavor and collider physics for many years to come.

\section*{Acknowledgments}
We are grateful to David Eriksson and Johan Rathsman for interesting and useful discussions.

\bibliographystyle{JHEP}
\bibliography{2hdm}

\end{document}